\documentclass[conference,a4paper]{IEEEtran}
\usepackage{enumerate}
\usepackage[top=0.8in, bottom=1.5in, left=0.6in, right=0.6in]{geometry}
\usepackage{epstopdf}
\usepackage{cite}
\usepackage{amscd}
\usepackage{dsfont}
\usepackage{latexsym}
\usepackage{array}

\usepackage{tikz}\usetikzlibrary{calc}
\usepackage{tabularx}
\usepackage{epsfig}
\usepackage{graphicx,subfigure}
\usepackage{amsmath,mathrsfs}
\usepackage{amsthm}
\usepackage{amssymb}  %used for the font style \mathbb{}
\usepackage{amsbsy}
\usepackage{color}
\usepackage{stfloats}
\usepackage{algorithm}
\usepackage{algpseudocode}
\usepackage{bm}
\usepackage{bbm}
\usepackage{pifont}
\usepackage{accents}
\usepackage{multirow}
\usepackage{booktabs}

\newtheorem{ther}{Theorem}
\newtheorem{deft}{Definition}

\theoremstyle{definition}
\newtheorem{rem}{Remark}

\DeclareMathOperator{\fit}{\mathrm{fit}}
\DeclareMathOperator{\thrs}{\mathrm{th}}
\ifCLASSINFOpdf
\else
\fi
\begin{document}
\title{On Polar Coding with Feedback}

\author{
\IEEEauthorblockN{Ling Liu$^1$, Qi Cao$^1$, Liping Li$^2$, Baoming Bai$^1$}
\IEEEauthorblockA{$^1$Guangzhou Institute of Technology, Xidian University, Guangzhou, China}
\IEEEauthorblockA{$^2$Key Laboratory of Intelligent Computing and Signal Processing Ministry of Education, Anhui
University, Hefei, China}
\IEEEauthorblockA{liuling@xidian.edu.cn, caoqi@xidian.edu.cn, liping\_li@ahu.edu.cn, bmbai@mail.xidian.edu.cn}
}


%Telephone: (800) 555--1212, Fax: (888) 555--1212}


\maketitle

\begin{abstract}
In this work, we investigate the performance of polar codes with the assistance of feedback in communication systems. Although it is well known that feedback does not improve the capacity of memoryless channels, we show that the finite length performance of polar codes can be significantly improved as feedback enables genie-aided decoding and allows more flexible thresholds for the polar coding construction. To analyze the performance under the new construction, we then propose an accurate characterization of the distribution of the error event under the genie-aided successive cancellation (SC) decoding. This characterization can be also used to predict the performance of the standard SC decoding of polar codes with rates close to capacity.  
\end{abstract}


% \fi
\IEEEpeerreviewmaketitle

\section{Introduction}
Feedback has become an increasingly indispensable resource in modern communication systems. For reliable communications, the receiver sends ``ACK/NACK '' signals back to the transmitter to decide whether a retransmission round is needed in TCP/IP. The more efficient hybrid automatic repeat request (HARQ) protocol tries to make use of the received data and some incremental redundancy to recover the message. With the assistance of the so-called 1-bit feedback, rateless codes \cite{Raptor06} or fountain codes \cite{MacKayFountain} were proposed for broadcast channels, which provide better rate compatibility than traditional forward error correction codes. For wireless communications, feedback is playing a crucial role as it conveys channel state information, which triggers the beamforming, precoding and a series of optimization schedules at the transmitter.

In this work, we focus on the design of channel coding in the presence of feedback. Although Shannon proved that using feedback does not improve the capacity of memoryless point-to-point channels, it certainly changes the trade-off between error probability and complexity for error correction. A celebrated example is the coding scheme proposed by Schalkwijk and Kailath \cite{SK1966}, named as the SK coding for additive white Gaussian noise (AWGN) channel. It achieves the capacity with doubly exponential decay in the error probability, which is better than the best random coding without feedback. Note that the SK coding was originally designed using noiseless feedback, and it has been subsequently extended to the noisy feedback and secure communication scenarios \cite{NoisyFB17, DaiBinFB20}.  

As a kind of capacity achieving codes, polar codes have received widespread attention since their invention in 2009. In the original work \cite{arikan2009channel}, Ar{\i}kan proposed the successive cancellation (SC) decoding algorithm, which is sub-optimal in finite block length yet still sufficient to show the capacity-achieving result. Since then, researchers have made great efforts to improve the performance of polar codes by more sophisticated decoding methods, such as belief propagation decoding, successive cancellation list (SCL) decoding \cite{ListPolar} and successive cancellation hybrid (SCH) decoding \cite{NiuKaiSCH2013}. Notably, with the assistance of cyclic redundancy check (CRC), polar codes under the CRC aided SCL decoding exhibit high competitiveness compared to LDPC and Turbo codes. More recently, Ar{\i}kan proposed the polarization adjusted convolutional (PAC) codes \cite{ArikanPAC}, which demonstrated the capability of achieving the dispersion approximation bound. An excellent survey on the development of polar codes can be found in the work \cite{NiuKaiGoldenPolar}. It appears that existing techniques for polar codes are close to reaching the performance limit in the absence of feedback. Perhaps it is time to introduce feedback to probe for further improvement potential. To the best of our knowledge, there is a limited body of literature on polar codes with feedback. In \cite{VakiFB15}, the authors proposed a scheme using a feedback link to return the potentially erroneous positions for short polar codes. A repetition code is then required to indicate the value of the querying bit. The optimized strategy for this scheme is still unknown. In this work, we provide a comprehensive characterization of the performance of polar codes under perfect feedback, including their coding rate, block error probability and decoding delay. Despite the assumption of perfect feedback, the analytical statistical model introduced in this paper will bring benefits to polar channel coding without feedback and polar lossless source coding, as can be seen in Sect. IV.

The rest of the paper is organized as follows: Sect. II gives preliminaries of polar codes and a brief review of the SK coding. The feedback polar coding scheme is described in Sect. III, where we also provide a detailed performance analysis of our coding scheme. In Sect. IV, we show the simulation results and how this analysis can be utilized for other related problems. The paper is concluded in Sect. V. 

$\it{Notation:}$ All random variables (RVs) are denoted by capital letters. Let $P(X)$ denote the probability mass function of a RV $X$ taking values in a countable set $\mathcal{X}$. The combination of $N$ i.i.d. copies of $X$ is denoted by a vector $X_1^{N}$ or $X^{[N]}$, where $[N]=\{1,...,N\}$, and its $i$-th element is given by $X_i$. The subvector of $X^{[N]}$ with indices limited to a subset $\mathcal{F} \subseteq [N]$ is denoted by $X^{\mathcal{F}}$. The cardinality of $\mathcal{F}$ is $|\mathcal{F}|$. The expectation and variance of a RV are denoted by $\mathsf{E}[\cdot]$ and $\mathsf{Var}[\cdot]$, respectively. We use binary logarithm $\log$ throughout this paper.

\section{Preliminaries of Polar Codes and the SK Coding Scheme}\label{sec:background}

\subsection{Polar Codes for Channel Coding}
Denote by $W$ a binary-input memoryless symmetric channel (BMSC) with uniform input $X \in \mathcal{X}$ and output $Y \in \mathcal{Y}$. Its channel transition probability is given by $P_{Y|X}$. The Shannon capacity of $W$ is denoted by $C(W)$. Let $N=2^n$ for some positive integer $n$. The polarization matrix is defined as
\begin{eqnarray}
\textbf{G}_N \triangleq \left[\begin{matrix}1&0\\1&1\end{matrix}\right]^{\otimes n} \times \textbf{B}_N,
\end{eqnarray}where $\otimes$ denotes the Kronecker product, and $\textbf{B}_N$ is the bit-reverse permutation matrix \cite{arikan2009channel}. The matrix $\textbf{G}_N$ transforms $N$ identical copies of $W$ into a vector channel $W_N: U^{[N]}\to Y^{[N]}$, where $U^{[N]}=X^{[N]}\textbf{G}_N^{-1}$ is the polarized version of the input $X^{[N]}$. Notice that $\textbf{G}_N^{-1}=\textbf{G}_N$ in $\mathbb{F}_2$. $W_N$ can be successively split into $N$ binary memoryless symmetric synthetic channels $U^i\to (U_1^{i-1}, Y^{[N]})$, denoted by $W_{N}^{(i)}$ with $1 \leq i \leq N$. The core of channel polarization states that, as $N$ increases, $W_N^{(i)}$ polarizes to a good (roughly error-free) channel or a totally noisy one almost surely. Moreover, the fractions of these two extreme synthetic channels turn to $C(W)$ and $1-C(W)$, respectively. To achieve the capacity, one can choose a rate $R=\frac{K}{N}$ smaller than $C(W)$, transmit information bits through $K$ good synthetic channels, and feed frozen bits (commonly set to all-zero) to the rest ones.

The indices of good or bad synthetic channels can be identified based on their associated Bhattacharyya parameters.
\begin{deft}\label{deft:symZ&asymZ}
Given a BMSC $W$ with transition probability $P_{Y|X}$, the Bhattacharyya parameter of $W$ is defined as
\begin{eqnarray}
Z(W)\triangleq\sum\limits_{y} \sqrt{P_{Y|X}(y|0)P_{Y|X}(y|1)}.
\end{eqnarray}
\end{deft}

For the $i$-th synthetic channel $W_N^{(i)}$, denote by $P_e(U^i|U_1^{i-1},Y^{[N]})$ the average error probability in estimating $U^i$ on the basis of $U_1^{i-1}$ and $Y^{[N]}$ via the maximum a posteriori probability (MAP) decision rule (see also \cite[eqn. (2.9)]{HassaniThesis}). For performance analysis, the Bhattacharyya parameter $Z(W_N^{(i)})$ provides an upper-bound on $P_e(U^i|U_1^{i-1},Y^{[N]})$. In \cite{arikan2009rate}, the rate of polarization $\beta$ is analyzed to characterize how fast $Z(W_N^{(i)})$ approaches 0 or 1. For the 2-by-2 kernel $\left[\begin{smallmatrix}1&0\\1&1\end{smallmatrix}\right]$, $\beta$ is upper-bounded by $\frac{1}{2}$. That is, for any $0<\beta<\frac{1}{2}$, we have
\begin{eqnarray}
\lim_{N\to\infty} P (Z(W_{N}^{(i)}) <2^{-N^{\beta}}) = C(W),
\end{eqnarray}and
\begin{eqnarray}
\lim_{N\to\infty} P (Z(W_{N}^{(i)}) >1-2^{-N^{\beta}}) = 1-C(W).
\end{eqnarray}
As a result, when the information set is chosen as $\mathcal{I}=\{i\in [N]:Z(W_{N}^{(i)})< 2^{-N^{\beta}}\}$ for channel coding, the block error probability under the SC decoding can be bounded as
\begin{eqnarray}
P_e^{SC}\leq N\cdot 2^{-N^{\beta}}.
\end{eqnarray}

\subsection{The SK Feedback Coding}
\begin{figure}[ht]
    \centering
    \includegraphics[width=4cm]{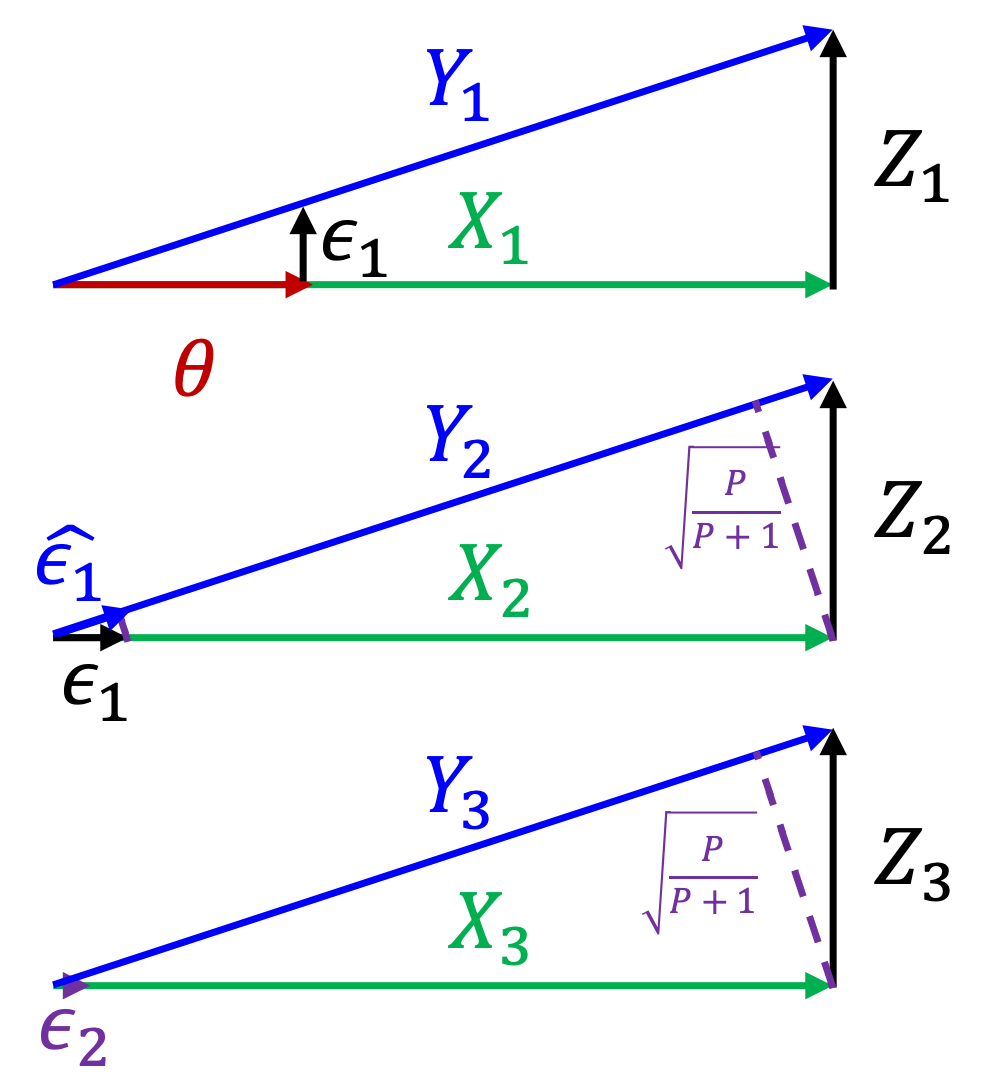}
    \caption{A geometric interpretation of the SK coding scheme.}
    \label{fig:SK}
\end{figure}

The SK feedback coding scheme was firstly proposed for the scalar AWGN channel with a noiseless and power-unconstrained feedback link. For initialization, the interval $[-\sqrt{3},\sqrt{3}]$ is divided into $M=2^{NR}$ isometric subintervals, the middle points of which correspond to the $M$ messages for transmission. Let $\theta$ denote the RV chosen from the set of middle points with probability $1/M$. One can check that $\mathsf{E}[\theta]=0$ and $\mathsf{E}[\theta^2]=1$ when $M$ is sufficiently large. For the 1st round, the transmitter sends symbol $X_1=\sqrt{P}\theta$, where $P$ is the power constraint. Then, $Y_1=X_1+Z_1$ is received, where $Z_1$ is the Gaussian noise RV with zero mean and unit variance. $Y_1$ is then perfectly fed back to the transmitter. For the 2nd round, the transmitter calculates $\epsilon_1 = \frac{Y_1-X_1}{\sqrt{P}}$ and sends $X_2=\frac{\sqrt{P}}{\sqrt{\mathsf{Var}[\epsilon_1]}}\epsilon_1$. After receiving $Y_2=X_2+Z_2$, the receiver performs minimum mean square error (MMSE) \cite{forney2003role} estimation for $\epsilon_1$ and obtains $\hat{\epsilon}_1$, which is then fed back for the next round. For the $i$-th round, where $i \geq 3$, let $\epsilon_{i-1}=\epsilon_{i-2}-\hat{\epsilon}_{i-2}$ and $X_i= \frac{\sqrt{P}}{\sqrt{\mathsf{Var}[\epsilon_{i-1}]}}\epsilon_{i-1}$. Based on $Y_i=X_i+Z_i$, the MMSE estimate $\hat{\epsilon}_{i-1}$ of $\epsilon_{i-1}$ is obtained and then fed back. This process repeats iteratively until $i=N$. For the recovery of $\theta$, notice that $\theta = \frac{X_1}{\sqrt{P}}$, which can be expanded as      
\begin{eqnarray}
\begin{aligned}
\frac{X_1}{\sqrt{P}} &= \frac{Y_1}{\sqrt{P}} - \epsilon_1 \\
&= \frac{Y_1}{\sqrt{P}} - \hat{\epsilon}_1 - (\epsilon_1-\hat{\epsilon}_1)\\
&= \underbrace{\frac{Y_1}{\sqrt{P}} - \sum_{i=1}^{N-1} \hat{\epsilon}_i}_{\hat{\theta}} - \epsilon_N ,
\end{aligned}
\end{eqnarray}
where $\hat{\theta}$ is treated as the estimate of $\theta$ since $Y_1$ and all $\hat{\epsilon}_i$'s are available at the receiver. An error event occurs when $\epsilon_N$ is larger than half of the subinterval size. By the property of MMSE, $\frac{\mathsf{Var}[\epsilon_i-\hat{\epsilon}_i]}{\mathsf{Var}[\epsilon_i]}=\frac{1}{P+1}$ for each step, which yields $\mathsf{Var}[\epsilon_N]=\frac{1}{P}(\frac{1}{P+1})^{N-1}$. Therefore, the error probability of the SK coding can be upper-bounded as
\begin{eqnarray}
P_e^{SK} \leq 2Q\left(\frac{\sqrt{3}/2^{NR}}{\sqrt{\mathsf{Var}[\epsilon_N]}}\right)=2Q(\delta \cdot 2^{N(C-R)}),
\end{eqnarray}
where $Q(\cdot)$ denotes the Q-function, $\delta$ is a constant depending on $P$, and $C$ is the capacity of the AWGN channel. A geometric interpretation of the first three rounds in the SK coding scheme is shown in Fig. \ref{fig:SK}, where the squared Euclidean distance represents the variance of a RV. 

\section{Polar Feedback Coding}\label{sec:PolarFB}
\subsection{The Coding Scheme}\label{sec:test}
The SK code provides a concrete example of how to use feedback to design a capacity-achieving code with good performance. However, the scalar form makes it incompatible with existing communication systems. We are more concerned with applying its underlying principles to improve the performance of existing block codes, particularly polar codes, which have been integrated into both 5G and future 6G standards. 

\begin{figure}[ht]
    \centering
    \includegraphics[width=8.5cm]{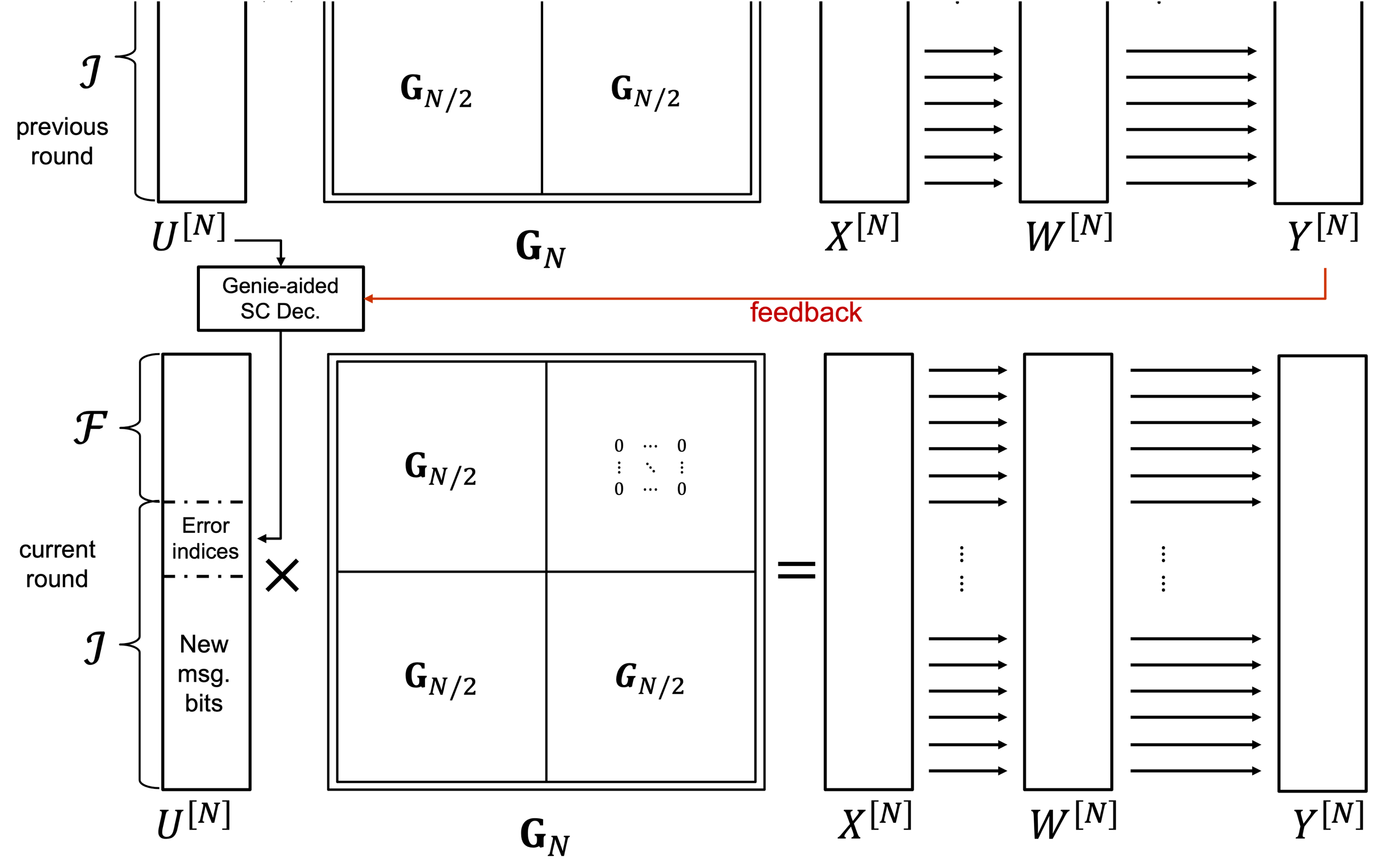}
    \caption{Block diagram of the polar feedback coding scheme.}
    \label{fig:FbkPolar}
\end{figure}

Our proposed polar feedback coding scheme is depicted in Fig. \ref{fig:FbkPolar}, where a series of polar coding blocks is chained by a feedback link. Suppose that a communication round has already been completed using standard polar coding. That is, the information bits and frozen bits are stitched into $U^{[N]}$, which is encoded to $X^{[N]}$ by $\mathbf{G}_N$. $X^{[N]}$ is then fed into $N$ i.i.d. copies of a BMSC $W$, producing output $Y^{[N]}$. Next, we assume that $Y^{[N]}$ is fed back to the transmitter before the following transmission round. Following the principle of the SK coding, one may perform a proper estimation of $U^{[N]}$ based on $Y^{[N]}$, and send the estimation error back to the receiver. Since $U^{[N]}$ is available at the transmitter, it turns out that the estimate can be obtained efficiently by a genie-aided (GA)-SC decoding algorithm. The estimation error is then represented by the following set of error-bit indices:
\begin{eqnarray}
\mathcal{T} \triangleq \{i\in \mathcal{I}: \hat{U}_i \neq U_i\},
\end{eqnarray}where $\hat{U}_i$ is the hard decision of the SC decoder based on $Y^{[N]}$ and its preceding bits $U_1^{i-1}$. We note that since the GA-SC decoder eliminates the error propagation effect of the standard SC decoder, the size $|\mathcal{T}|$ is significantly smaller than the case when $U^{[N]}$ is not available, as shown in \cite{BurgFlip14}. Each index in $\mathcal{T}$ can be described by a $\log N$-bit sequence and the estimation error requires $|\mathcal{T}|\times\log N$ bits. Unlike the SK coding, the next transmission is not solely dedicated to convey the estimation error. Instead, we append $K-|\mathcal{T}|\times\log N$ new information bits to maintain the dimension of polar codes, which is more convenient for practical implementation. To distinguish between the error indices and the new information bits, the size $|\mathcal{T}|$ should also be recorded, which requires at most $\log N$ bits, since $|\mathcal{T}| < N$. 
By doing so, we have assembled $U^{[N]}$ for the next coding block. At the receiver, standard SC decoding is utilized to test whether $U^{[N]}$ can be recovered from $Y^{[N]}$ at each round with the aid of CRC. The rate loss caused by the size $|\mathcal{T}|$ for truncation and CRC is ignored for simplicity. Once it succeeds, the previous coding blocks can be decoded in a reversed order when $\mathcal{T}$ is available. The delay $D$ ($D\geq 1$) of decoding is defined such that the $j$-th block is successfully decoded by the time the $(j+D-1)$-th block is received. This process proceeds iteratively until all information bits have been successively decoded. We temporarily assume that the buffer is of adequate size at the receiver side, meaning that the maximum delay tolerance $D_{\max}=\infty$. Let $\bar{D}$ denote the expected delay across different blocks.

\begin{rem}
Using a polar decoding algorithm for error estimation indeed endows the feedback link with a certain degree of noise tolerance. In practice, $Y^{[N]}$ can be converted into log-likelihood ratio with a certain level of quantization precision. As long as the decoding algorithm is synchronized between the two sides, a noisy version of $Y^{[N]}$ is also acceptable for our scheme. For convenience of description, however, we simply assume that $Y^{[N]}$ is perfectly returned.
\end{rem}

\begin{ther}\label{thm:FbRate}
For the proposed ideal feedback model, the optimal construction of polar codes in terms of coding rate is given by:
\begin{eqnarray}
\mathcal{F} \triangleq \left\{i\in [N]: P_e(U_i|U_1^{i-1},Y^{[N]}) > \frac{1}{\log N}\right\},
\end{eqnarray}and $\mathcal{I}=\mathcal{F}^c$, where $\frac{1}{\log N}$ is called the optimal threshold $\epsilon^*_{\thrs}$.
\end{ther}

\begin{IEEEproof}
If $P_e(U_i|U_1^{i-1},Y^{[N]})>\frac{1}{\log N}$, putting $i$ in the frozen set $\mathcal{F}$ costs 1-bit rate loss. However, putting $i$ in $\mathcal{I}$ costs average rate loss
\begin{eqnarray}
P_e(U_i|U_1^{i-1},Y^{[N]})\cdot \log N  > 1 
\end{eqnarray}for the next round. For the first round, no error index exists, which can be viewed as a closed-loop coding scheme that always starts with $\mathcal{T}=\emptyset$. From a long-term perspective, the influence of the initial state on the coding rate is negligible. 
\end{IEEEproof}

\begin{rem}
Compared with the construction of polar codes without feedback in Section \ref{sec:background}, where the threshold is chosen as $2^{-N^\beta}$, we see that feedback helps to relax the threshold to $\frac{1}{\log N}$, which results in a higher coding rate for finite block length. Note that the indices in $\mathcal{T}$ are not uniformly distributed. Combined with lossless compression techniques, each element can be represented with an average length of less than $\log N$. As a result, $\epsilon^*_{\thrs}$ can be further optimized. More details can be found in reference \cite[Sect. III-C]{LingCmprITW25}. In the rest of this paper, we still assume that each index is represented by $\log N$ bits.
\end{rem}

As a validation of Theorem \ref{thm:FbRate}, we present the Monte Carlo simulation results for our coding scheme with different thresholds $\epsilon_{\thrs}$ in Table \ref{tab:FbRate}, where $W$ is a binary symmetric channel (BSC) with crossover probability $0.11$. It can also be observed that the average delay varies dramatically with $\epsilon_{\thrs}$. Under the optimal threshold $\epsilon^*_{\thrs}$, the average delay exceeds 10; moreover, the simulated maximum delay can be over 100, which is impractical for real-world applications. In the next section, we investigate the relationship between $D$ and $\epsilon_{\thrs}$ and show that the performance of our scheme can be precisely predicted.

\begin{table}[htbp]
    \centering
    \caption{ Performance of polar codes with $N=2^{10}$ and different thresholds over BSC(0.11).}
    \label{tab:FbRate}
    \renewcommand{\arraystretch}{1.2}
    \setlength{\tabcolsep}{5pt}
    \begin{tabular}{l|cccccc} 
        \toprule 
        Threshold $\epsilon_{\thrs}$ & $\frac{1}{3}\epsilon^*_{\thrs}$ & $\frac{1}{2}\epsilon^*_{\thrs}$ & $\frac{1}{1.5}\epsilon^*_{\thrs}$ & $\epsilon^*_{\thrs}$ & $\frac{1}{0.8}\epsilon^*_{\thrs}$ & $\frac{1}{0.5}\epsilon^*_{\thrs}$ \\
        \midrule 
        Average rate & 0.407 & 0.416 & 0.422 & \textbf{0.426} & 0.424 & 0.407 \\
        Average delay $\bar{D}$ & 2.168 & 3.102 & 4.879 & 10.340 & 22.847 & 131.933 \\
        \bottomrule 
    \end{tabular}
\end{table}

\subsection{Performance Analysis}\label{sec:DistT}
It is clear that the delay $D$ follows a geometric distribution with successive probability $P(\hat{U}^{\mathcal{I}}=U^{\mathcal{I}})$ and $\bar{D}$ is equal to its reciprocal. Evaluating $P(\hat{U}^{\mathcal{I}}=U^{\mathcal{I}})$ is equivalent to evaluating $P_e^{SC}$. The union bound $\sum_{i\in\mathcal{I}}P_e(U_i|U_1^{i-1},Y^{[N]})$ mentioned in Section \ref{sec:background} generally provides a tight upper bound on $P_e^{SC}$ when $\epsilon_{\thrs}$ is close to 0. However, when $\epsilon_{\thrs}=O(\frac{1}{\log N})$, this union bound could be useless as it can exceed 1. To better predict the performance of this scheme, we turn to investigate the distribution of the size $|\mathcal{T}|$, which can be characterized by only a few parameters.

Given that $|\mathcal{T}|=\sum_{i \in \mathcal{I}} \mathbbm{1}(\hat{U}_i\neq U_i)$ with $\mathbbm{1}(\cdot)$ being the indicator function, our first observation is that $\mathsf{E}[|\mathcal{T}|]$ can be obtained as follows.
\begin{eqnarray}\label{eqn:ETC}
\mathsf{E}[|\mathcal{T}|]=\sum_{i \in \mathcal{I}} \mathsf{E}[\mathbbm{1}(\hat{U}_i\neq U_i)] = \sum_{i\in\mathcal{I}}P_e(U_i|U_1^{i-1},Y^{[N]}),
\end{eqnarray}where $P_e(U_i|U_1^{i-1},Y^{[N]})$ can be estimated within acceptable precision using the methods in \cite{IdoConstruct}. 

\begin{rem}
It is noted that $P(\hat{U}^{\mathcal{I}} \neq U^{\mathcal{I}})=P(|\mathcal{T}| \neq 0)$. Using the Markov inequality, we have
\begin{eqnarray}
P(|\mathcal{T}|\geq 1) \leq \frac{\mathsf{E}[|\mathcal{T}|]}{1}=\sum_{i\in\mathcal{I}}P_e(U_i|U_1^{i-1},Y^{[N]}),
\end{eqnarray}which is a restatement of the union bound for $P_e^{SC}$, and it may explain why this is not tight when $\mathsf{E}[|\mathcal{T}|]$ is not near 0.
\end{rem}

We then investigate the variance $\mathsf{Var}[|\mathcal{T}|]$ of $|\mathcal{T}|$. To do so, the second moment $\mathsf{E}[|\mathcal{T}|^2]$ is expressed as 
\begin{eqnarray}
\begin{aligned}
\mathsf{E}[|\mathcal{T}|^2] &= \mathsf{E}\left[\sum_{i \in \mathcal{I}} \mathbbm{1}(\hat{U}_i\neq U_i)\sum_{j \in \mathcal{I}} \mathbbm{1}(\hat{U}_j\neq U_j)\right]\\
&= \mathsf{E}\left[\sum_{i \in \mathcal{I}} \mathbbm{1}(\hat{U}_i\neq U_i)\sum_{j=i} \mathbbm{1}(\hat{U}_j\neq U_j)\right] \\
&\;\;\;+ \mathsf{E}\left[\sum_{i \in \mathcal{I}} \mathbbm{1}(\hat{U}_i\neq U_i)\sum_{j \neq i, j\in \mathcal{R}(i)} \mathbbm{1}(\hat{U}_j\neq U_j)\right]\\
&\;\;\;+ \mathsf{E}\left[\sum_{i \in \mathcal{I}} \mathbbm{1}(\hat{U}_i\neq U_i)\sum_{j\neq i, j\in \mathcal{R}^c(i)} \mathbbm{1}(\hat{U}_j\neq U_j)\right], 
\end{aligned}
\end{eqnarray}where $\mathcal{R}(i)$ is defined as the error indices that are correlated with the index $i$. The first term on the r.h.s. of the above equality is $\mathsf{E}[|\mathcal{T}|]$. For the third term, based on the results in \cite{AlsanBEC25} and \cite{ParizeBEC13}, we claim that the majority of the error indices tend to be uncorrelated with each other. Moreover, when $\epsilon_{\thrs}$ is chosen as $O(\frac{1}{\log N})$, the product of the expectations of two error events is $O(\frac{1}{\log^2 N})$, which is insignificant compared to $\mathsf{E}[|\mathcal{T}|]$. Therefore, the third term can be approximated by $\mathsf{E}^2[|\mathcal{T}|]$. Rearranging the equation yields the following approximation for the second term.
\begin{eqnarray}
\begin{aligned}
\mathsf{Var}[|\mathcal{T}|]&-\mathsf{E}[|\mathcal{T}|] \\
&\approx\mathsf{E}\left[\sum_{i \in \mathcal{I}} \mathbbm{1}(\hat{U}_i\neq U_i)\sum_{j \neq i, j\in \mathcal{R}(i)} \mathbbm{1}(\hat{U}_j\neq U_j)\right].
\end{aligned}
\end{eqnarray}

%\begin{eqnarray}
%\begin{aligned}
%\mathsf{Var}[|\mathcal{T}|]&-\mathsf{E}[|\mathcal{T}|] \\
%&\approx\mathsf{E}\Bigg[\sum_{i \in \mathcal{I}} \mathbbm{1}(\hat{U}_i\neq U_i)\underbrace{\sum_{j \neq i, j\in \mathcal{R}(i)} \mathbbm{1}(\hat{U}_j\neq %U_j)}_{\frac{\mathsf{E}[|\mathcal{T}|]}{r_{\fit}}}\Bigg].
%\end{aligned}
%\end{eqnarray}

\begin{figure}[ht]
    \centering
    \includegraphics[width=8cm]{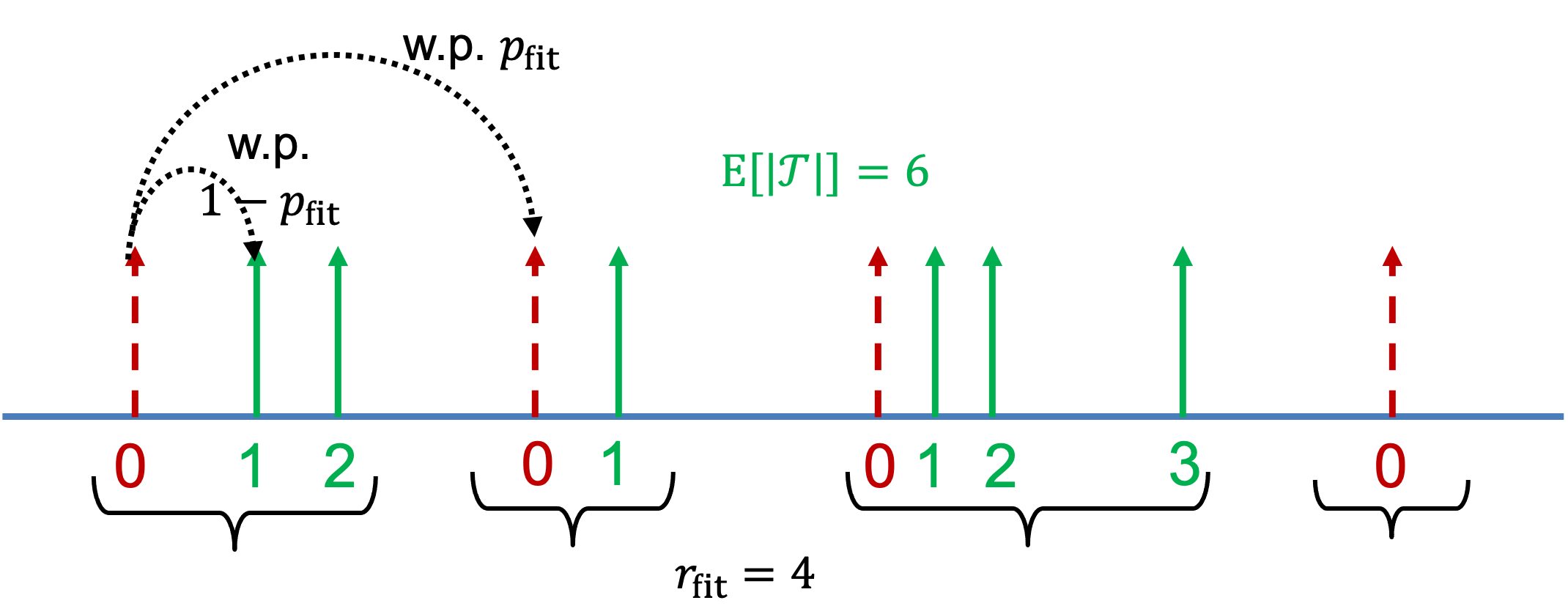}
    \caption{A statistic perspective of the occurrence of error events.}
    \label{fig:NBDemo}
\end{figure}

Characterizing the exact pattern of $\mathcal{R}(i)$ is nontrivial. Alternatively, we adopt an ``amortized'' approach to view the occurrence of error event. The key idea is to uniformly distribute error events across $r_{\fit}$ disjoint subsets of information indices, with the assumption that correlations are confined within each subset. That is, the second term is decoupled by letting $\sum_{j \neq i, j\in \mathcal{R}(i)} \mathbbm{1}(\hat{U}_j\neq U_j)=\frac{\mathsf{E}[|\mathcal{T}|]}{r_{\fit}}$ for some positive $r_{\fit}$. Since all $\mathsf{E}[|\mathcal{T}|]$ errors are assigned to some indices in $\mathcal{R}(i)$ for different $i$, the index $i$ corresponding to $\mathbbm{1}(\hat{U}_i\neq U_i)$ can be viewed as a switching error index and the indices in $\mathcal{R}(i)$ are triggered by such $i$. Noting that $\mathsf{E}[\sum_{i \in \mathcal{I}} \mathbbm{1}(\hat{U}_i\neq U_i)]$ equals $\mathsf{E}[|\mathcal{T}|]$, there are $r_{\fit}$ such switching indices. The error occurrence in $\mathcal{R}(i)$ is then modeled as a geometric process (starts with index 0) with successive probability $p_{\fit}$, which captures the possibility that the next error is a switching error or a triggered error. A toy example with $r_{\fit}=4$ and $\mathsf{E}[|\mathcal{T}|]=6$ is demonstrated in Fig. \ref{fig:NBDemo}, with switching error indices indicated by red dashed arrows and triggered error indices by green solid ones. The index of the geometric process is also shown for each of the $r_{\fit}$ trials. Note that $r_{\fit}$ and $\mathsf{E}[|\mathcal{T}|]$ are not necessarily integers; they can also be positive real numbers. 

The above analysis suggests that it is possible to view the RV $|\mathcal{T}|$ as the result of $r_{\fit}$ consecutive geometric processes with successive probability $p_{\fit}$, which is indeed a \emph{negative binomial} process with number of successes $r_{\fit}$. 

\begin{ther}\label{thm:NBfit}
Given $\mathsf{E}[|\mathcal{T}|]$ and $\mathsf{Var}[|\mathcal{T}|]$, suppose that $X$ is a RV that follows the negative binomial distribution with $r_{\fit}=\frac{\mathsf{E}^2[|\mathcal{T}|]}{\mathsf{Var}[|\mathcal{T}|]-\mathsf{E}[|\mathcal{T}|]}$ and $p_{\fit}=\frac{\mathsf{E}[|\mathcal{T}|]}{\mathsf{Var}[|\mathcal{T}|]}$, i.e., for  $x=0,1,2,...,$
\begin{eqnarray}
P_{\mathrm{NB}}(X=x|r_{\fit},p_{\fit})=\binom{r_{\fit}+x-1}{x}p^{r_{\fit}}_{\fit}(1-p_{\fit})^x, 
\end{eqnarray}where the binomial coefficient can be expressed equivalently using Euler's Gamma function for real $r_{\fit}$. Then, $\mathsf{E}[X]=\mathsf{E}[|\mathcal{T}|]$ and $\mathsf{Var}[X]=\mathsf{Var}[|\mathcal{T}|]$.
\end{ther}
\begin{IEEEproof}
Note that the negative binomial process is made up of $r_{\fit}$ identical geometric processes. The expectation and variance of the geometric distributions (starts with index 0) with successive probability $p_{\fit}$ are $\frac{1-p_{\fit}}{p_{\fit}}$ and $\frac{1-p_{\fit}}{p^2_{\fit}}$, respectively.
\end{IEEEproof}

\begin{figure}[ht]
    \centering
    \includegraphics[width=8cm]{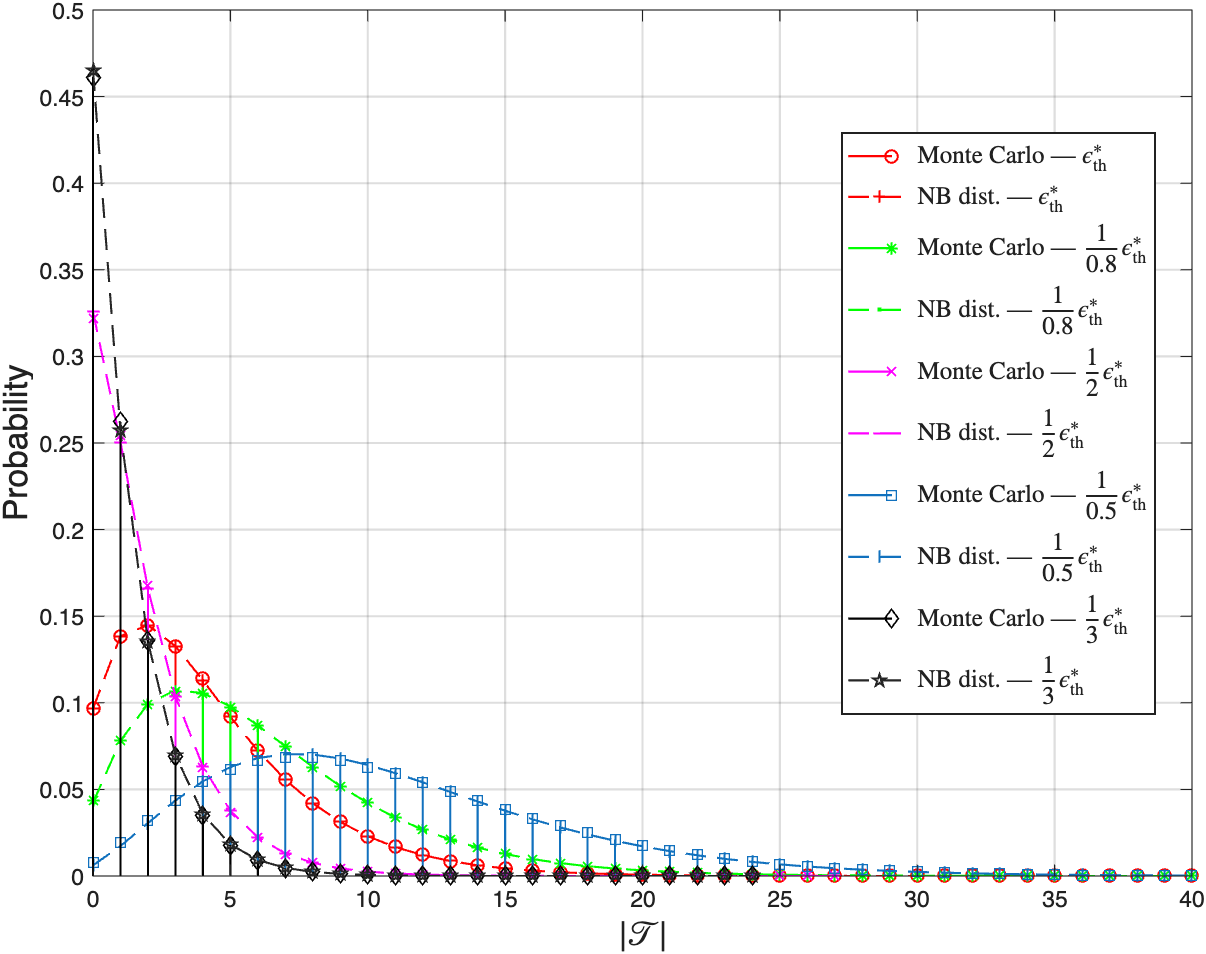}
    \caption{The fitness of the negative binomial distribution for $|\mathcal{T}|$.}
    \label{fig:NBFit}
\end{figure}

The fitness of the negative binomial distribution for $|\mathcal{T}|$ with varying $\epsilon_{\thrs}$ for the same environment as in Table \ref{tab:FbRate} is illustrated in Fig. \ref{fig:NBFit}. Similar results can be obtained when the channel is replaced with a binary erasure channel (BEC) or a binary input AWGN (BIAWGN) channel with BPSK modulation. We see that Theorem \ref{thm:NBfit} provides a good prediction of the distribution of $|\mathcal{T}|$. Consequently, for a specific $\epsilon_{\thrs}$, once $\mathsf{E}[|\mathcal{T}|]$ and $\mathsf{Var}[|\mathcal{T}|]$ are known, the successive decoding probability for each round in our feedback coding scheme can be approximated by $P(|\mathcal{T}|=0)=p_{\fit}^{r_{\fit}}$. Then, the delay $D$ follows a geometric distribution with successive probability $p_{\fit}^{r_{\fit}}$ and $\bar{D}=p_{\fit}^{-r_{\fit}}$. For a finite maximum delay tolerance $D_{\max}$, the decoding failure probability can be also predicted by the sum of the tails of the geometric distribution, which is $(1-p_{\fit}^{r_{\fit}})^{D_{\max}}$. Recall that $\mathsf{E}[|\mathcal{T}|]$ can be obtained from \eqref{eqn:ETC}. Next, we will show that $\mathsf{Var}[|\mathcal{T}|]$ may also be accurately estimated by investigating the second order statistical properties of the channel polarization process. 

Let $W$ be a BEC($p$) with erasure probability $p$. Consider the polarization process of $W$ with $N=2^n$. To save space, we derive the next theorem based on the result of \cite{ParizeBEC13}. Before that, we define the erasure event for index $i$ as $\mathcal{E}_i \triangleq \mathbbm{1}(\hat{U}_i=?)$. Since $W_N^{(i)}$ is a BEC for all $i \in [N]$, $\mathsf{E}[\mathcal{E}_i]=Z(W_N^{(i)})$. Denote by $\mathbf{C}_N$ the $N$-by-$N$ covariance matrix of the random vector consisting of $\mathcal{E}_i$ for $i\in [N]$, i.e.,
\begin{eqnarray}
\mathbf{C}_N(i,j) \triangleq \mathsf{Cov}[\mathcal{E}_i, \mathcal{E}_j],
\end{eqnarray}where $\mathsf{Cov}[X,Y] \triangleq \mathsf{E}[XY]-\mathsf{E}[X]\mathsf{E}[Y]$. 

%For better presentation, an index $i$ is interchangeably expressed by its binary expansion $\mathbf{s} \in\{0,1\}^n.$

\begin{ther}\label{thm:BECVarT}
Consider the polar feedback coding when $W$ is a BEC($p$). The Bhattacharyya parameter $Z(W_N^{(i)})$ can be calculated by applying the transform $(p,p) \to (2p-p^2,p^2)$ recursively. $\mathbf{C}_N$ can also be calculated based on $\mathbf{C}_{N/2}$ as in \eqref{eqn:CovBEC} with initial condition $\mathbf{C}_1=p(1-p)$. Given a threshold $\epsilon_{\thrs}$ for $P_e(U_i|U_1^{i-1},Y^{N})$ and constructing $\mathcal{I}$ as in Theorem \ref{thm:FbRate}, then, 
\begin{eqnarray}
\mathsf{E}[|\mathcal{T}|]= \sum_{i \in \mathcal{I}} P_e(U_i|U_1^{i-1},Y^{N}) =\frac{1}{2}\sum_{i \in \mathcal{I}}Z(W_N^{(i)}),   
\end{eqnarray}and
\begin{eqnarray}
\mathsf{Var}[|\mathcal{T}|]= \frac{1}{2}\mathsf{E}[|\mathcal{T}|]+\frac{1}{4}\sum_{i \in \mathcal{I}}\sum_{j \in \mathcal{I}} \mathbf{C}_N(i,j).   
\end{eqnarray}
\end{ther}

\begin{IEEEproof}
The first part is straightforward as $P_e(W)=\frac{1}{2}Z(W)$ for BEC. For the second part, \eqref{eqn:CovBEC} is from \cite{ParizeBEC13}. The proof is completed by expanding $\mathsf{Var}[|\mathcal{T}|]$ as \eqref{eqn:VarTCalc}, where the third equality holds because $\mathsf{E}[\mathbbm{1}(\hat{U}_i\neq U_i)]=\frac{1}{2}\mathsf{E}[\mathbbm{1}(\hat{U}_i=?)]$ and $\mathsf{E}[\mathbbm{1}(\hat{U}_i\neq U_i)\mathbbm{1}(\hat{U}_j\neq U_j)]=\frac{1}{4}\mathsf{E}[\mathbbm{1}(\hat{U}_i=?)\mathbbm{1}(\hat{U}_j=?)]$ for $i\neq j$ and uniform $U^{\mathcal{I}}$.
\end{IEEEproof}

\begin{figure}[ht]
    \centering
    \includegraphics[width=8cm]{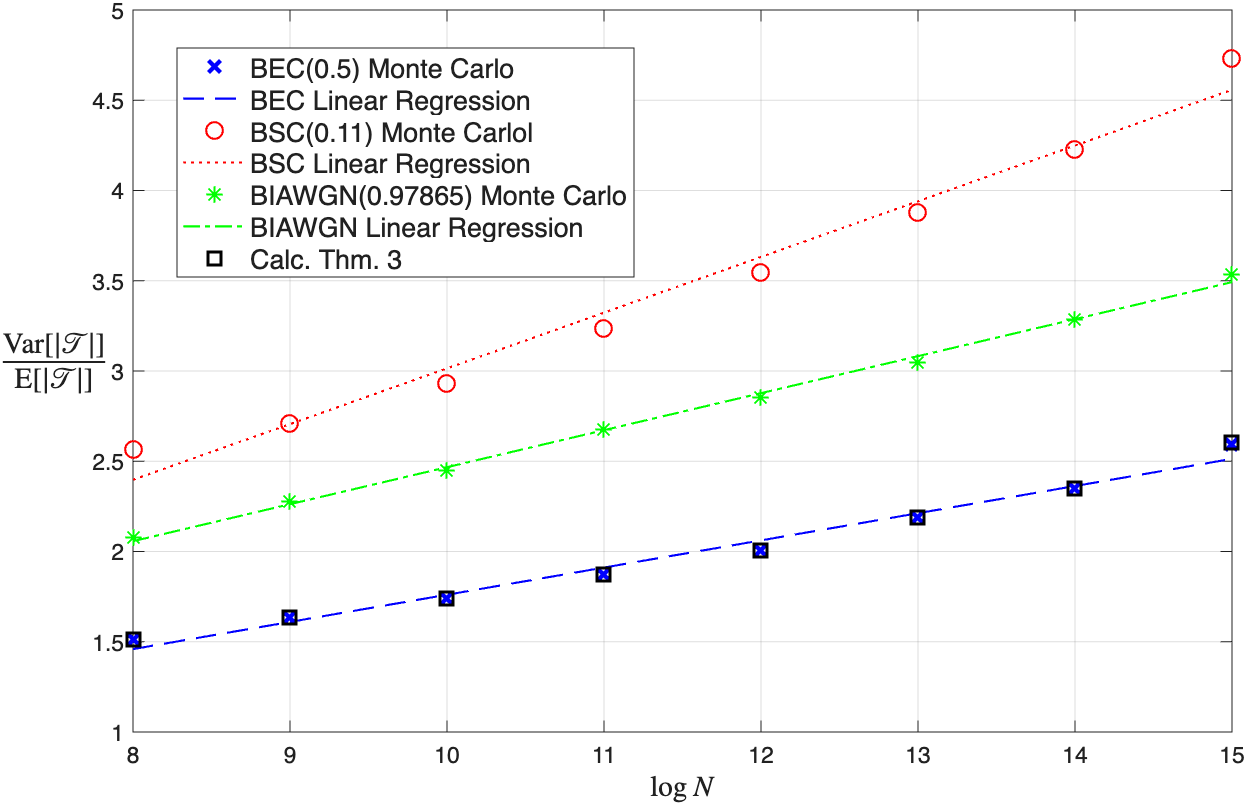}
    \caption{$\frac{\mathsf{Var}[|\mathcal{T}|]}{\mathsf{E}[|\mathcal{T}|]}$ for different BMSCs and $N$.}
    \label{fig:VoE}
\end{figure}

Fig. \ref{fig:VoE} demonstrates the validity of Theorem \ref{thm:BECVarT} over BEC(0.5) with varying $N$ and $\epsilon^*_{\thrs}=\frac{1}{\log N}$. We also plot the results for BSC(0.11) and BIAWGN channel with standard deviation $\sigma=0.97865$. An approximately linear relationship between $\frac{\mathsf{Var}[|\mathcal{T}|]}{\mathsf{E}[|\mathcal{T}|]}$ and $\log N$ can be observed, at least for $N$ from $2^9$ to $2^{14}$, which means it is possible to estimate $\mathsf{Var}[|\mathcal{T}|]$ based on $\mathsf{E}[|\mathcal{T}|]$ for channels other than BEC. We may also use the BEC approximation \cite{arkan2008performance} or the channel merging methods \cite{PolarConstru, IdoConstruct} to evaluate $\mathsf{Var}[|\mathcal{T}|]$ directly for future work.

\section{Simulation Results}\label{sec:Simu}
\begin{figure}[ht]
    \centering
    \includegraphics[width=8cm]{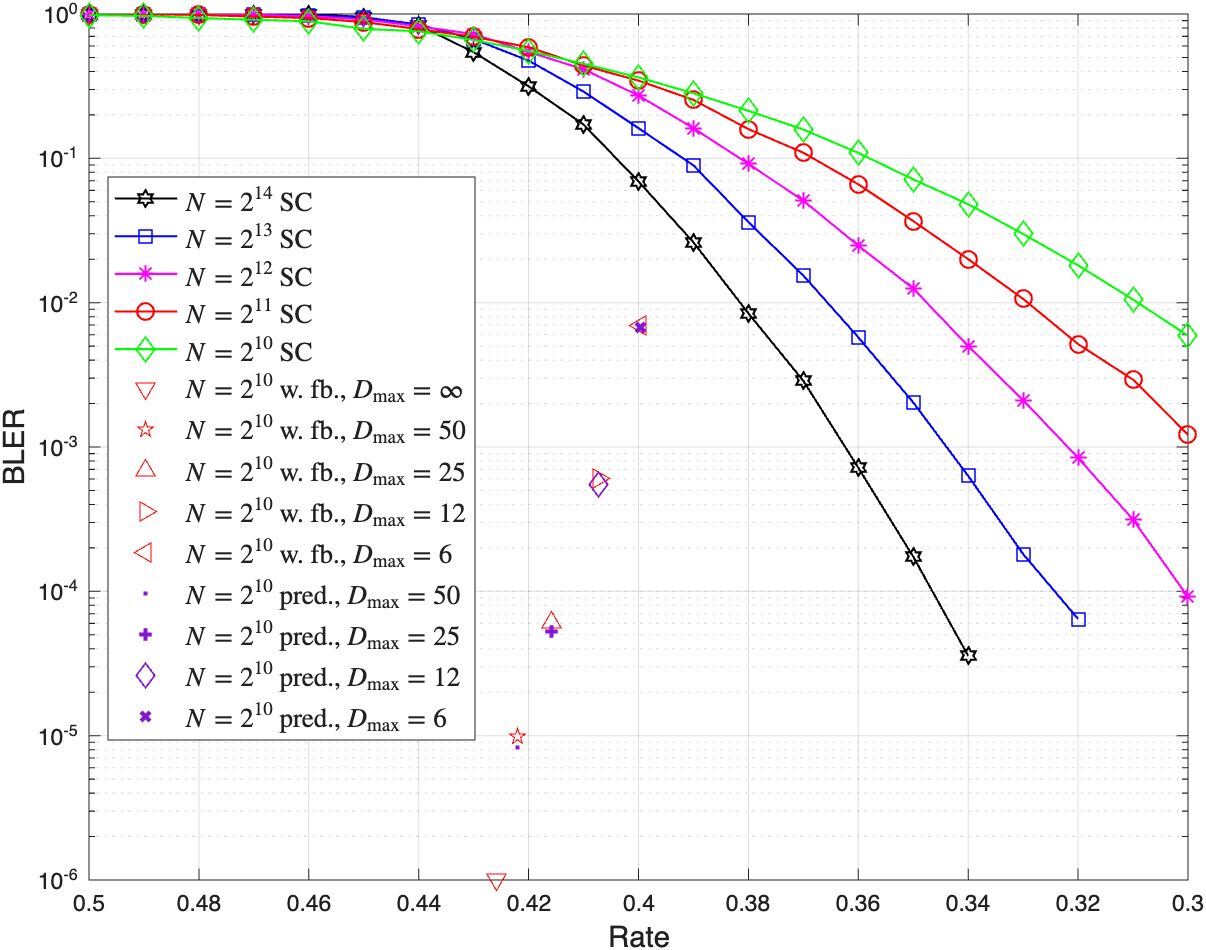}
    \caption{The error performance of polar feedback coding for BSC(0.11).}
    \label{fig:BSCFB}
\end{figure}

The block error rate (BLER) performance of our proposed feedback coding scheme for BSC(0.11) is shown in Fig. \ref{fig:BSCFB}. For comparison, the performance curves without feedback are also provided to demonstrate the improvement. When $D_{\max} = \infty$, the system attains theoretical error-free performance. This point is marked at a BLER of $10^{-6}$ in the figure. For other points, we adjust the threshold $\epsilon_{\thrs}$ as in Table \ref{tab:FbRate} to control the distribution of $|\mathcal{T}|$ and $D$ for different tolerance $D_{\max}$. These points characterize the trade-off between rate and feedback delay. Thanks to Theorem \ref{thm:NBfit}, $P(|\mathcal{T}|)$ can be accurately estimated, which, in turn, enables precise prediction of the BLER.

\begin{figure}[ht]
    \centering
    \includegraphics[width=8cm]{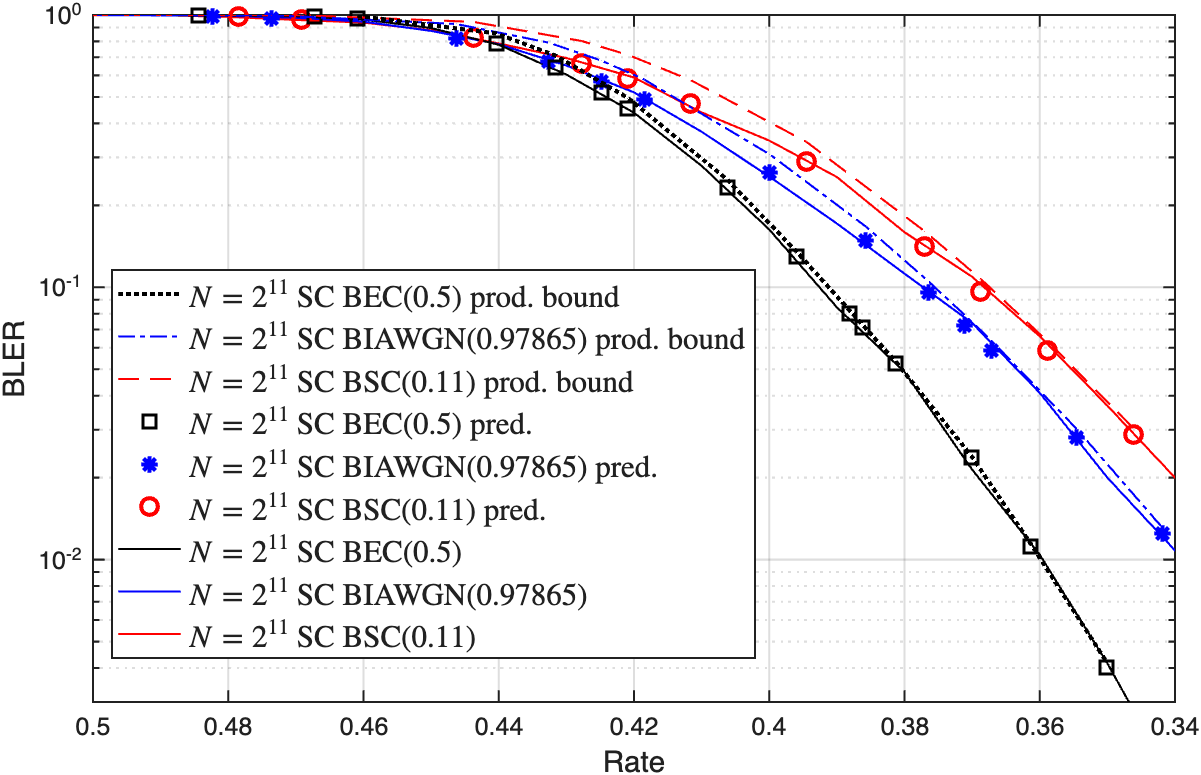}
    \caption{Predicting the BLER of the standard SC decoding using Theorem \ref{thm:NBfit}.}
    \label{fig:BFit}
\end{figure}

As a by-product, the approximation for $P(|\mathcal{T}|=0)$ according to Theorem \ref{thm:NBfit} provides a means to predict the BLER of the standard SC decoder (without feedback). Note that the rate is now $\frac{|\mathcal{I}|}{N}$ instead of $\frac{|\mathcal{I}|-n\mathsf{E}[|\mathcal{T}|] }{N}$. In Fig. \ref{fig:BFit}, the block length is set to $N=2^{11}$ for three typical BMSCs: BEC(0.5), BSC(0.11) and BIAWGN(0.97865). By adjusting the threshold as $\epsilon_{\thrs}=\frac{1}{\alpha}\epsilon^*_{\thrs}$ for $\alpha$ from $0.5$ to $200$, we see that the performance of SC decoding can be well predicted by Theorem \ref{thm:NBfit} using $\mathsf{E}[|\mathcal{T}|]$ and $\mathsf{Var}[|\mathcal{T}|]$, which may be computable during the construction of polar codes. Combining our method with the existing tight bounds \cite{IdoConstruct,arikan2009rate,ParizeBEC13} for lower BLER region, the entire performance curves can be predicted. For comparison, we also plot the product bound $1-\prod_{i\in\mathcal{I}}(1-P_e(U^i|U_1^{i-1},Y^{[N]}))$ in Fig. \ref{fig:BFit}, which assumes that the error events are independent of each other. As can be seen, Theorem \ref{thm:NBfit} provides more accurate prediction by taking the correlation between error events into consideration, especially in the high BLER region. 

As another by-product, we show that Theorem \ref{thm:NBfit} can be employed to compress the number $|\mathcal{T}|$, when the polar lossless coding scheme \cite{cronie2010lossless} is operating in a serial data-stream mode, where $|\mathcal{T}|$ is required for truncation. For the i.i.d. Bernoulli (BER) source with probability 0.11 and length $N=2^{10}$, we derive an empirical estimate of $P(|\mathcal{T}|)$ via Monte Carlo simulation for different thresholds. The entropy of $P(|\mathcal{T}|)$ is denoted by $H(|\mathcal{T}|)$. We then use $P_{\mathrm{NB}}(|\mathcal{T}|)$ from Theorem \ref{thm:NBfit}, with entropy $H_{\mathrm{NB}}(|\mathcal{T}|)$, to generate a Huffman dictionary for $|\mathcal{T}|$ and compute its average length $\bar{L}$. $H(|\mathcal{T}|)$, $H_{\mathrm{NB}}(|\mathcal{T}|)$ and $\bar{L}$ (all in bits) are listed in Table \ref{tab:CprT}, which shows the validity of our statistical model.

\begin{table}[htbp]
    \centering
    \caption{The compression of $|\mathcal{T}|$ for Ber(0.11) and $N=2^{10}$, using the distribution in Theorem \ref{thm:NBfit}.}
    \label{tab:CprT}
    \renewcommand{\arraystretch}{1.2}
    \setlength{\tabcolsep}{5pt}
    \begin{tabular}{l|cccccc} 
        \toprule 
        $\epsilon_{\thrs}$ & $\frac{1}{3}\epsilon^*_{\thrs}$ & $\frac{1}{2}\epsilon^*_{\thrs}$ & $\frac{1}{1.5}\epsilon^*_{\thrs}$ & $\epsilon^*_{\thrs}$ & $\frac{1}{0.8}\epsilon^*_{\thrs}$ & $\frac{1}{0.5}\epsilon^*_{\thrs}$ \\
        \midrule 
        $H({|\mathcal{T}|})$    & 2.0978 & 2.5739 & 3.0224 & 3.5746 & 3.9960 & 4.5965 \\
        $H_{\mathrm{NB}}(|\mathcal{T}|)$ & 2.0983 & 2.5751 & 3.0240 & 3.5748 & 3.9928 & 4.5844 \\
        $\bar{L}$               & 2.1026 & 2.6248 & 3.0611 & 3.6002 & 4.0369 & 4.6164 \\
        \bottomrule 
    \end{tabular}
\end{table}

\section{Conclusion}
In this work, we investigate the performance of polar codes with feedback. Specifically, we develop a statistical method to characterize the number of error events under the GA-SC decoder, which captures the interplay among coding rate, system delay, and error performance. As a direction for future work, this statistical method can be extended to characterize systems with a limited error index budget—specifically, to analyze the GA-$T_{\max}$ SC decoder where $T_{\max}$ is the maximum allowable genie corrections. The coding rate can be further improved by compressing $\mathcal{T}$ using our recent work \cite{LingCmprITW25}. We also believe that the framework developed here can be applied to advance the finite-length performance analysis \cite{HassaniScal14,MondelliScal16} of polar codes. 
\begin{figure*}[th]
\begin{eqnarray}\label{eqn:CovBEC}
\begin{aligned}
\mathbf{C}_N(i,j) = 
\begin{cases}
2 \left(1-Z\Big(W_{N/2}^{(\frac{i+1}{2})}\Big)\right) \left(1-Z\Big(W_{N/2}^{(\frac{j+1}{2})}\Big)\right) \mathbf{C}_{N/2}(\frac{i+1}{2},\frac{j+1}{2})+\mathbf{C}_{N/2}(\frac{i+1}{2},\frac{j+1}{2})^2, &\text{ if $i$ is odd and $j$ is odd} \\
2  \left(1-Z\Big(W_{N/2}^{(\frac{i+1}{2})}\Big)\right) Z\Big(W_{N/2}^{(\frac{j+1}{2})}\Big)\mathbf{C}_{N/2}(\frac{i+1}{2},\frac{j}{2})-\mathbf{C}_{N/2}(\frac{i+1}{2},\frac{j}{2})^2, &\text{ if $i$ is odd and $j$ is even} \\
2  Z\Big(W_{N/2}^{(\frac{i}{2})}\Big) \left(1-Z\Big(W_{N/2}^{(\frac{j+1}{2})}\Big)\right)\mathbf{C}_{N/2}(\frac{i}{2},\frac{j+1}{2})-\mathbf{C}_{N/2}(\frac{i}{2},\frac{j+1}{2})^2, &\text{ if $i$ is even and $j$ is odd} \\
2  Z\Big(W_{N/2}^{(\frac{i}{2})}\Big) Z\Big(W_{N/2}^{(\frac{j}{2})}\Big)\mathbf{C}_{N/2}(\frac{i}{2},\frac{j}{2})-\mathbf{C}_{N/2}(\frac{i}{2},\frac{j}{2})^2, &\text{ if $i$ is even and $j$ is even}
\end{cases}
\end{aligned}
\end{eqnarray}
\hrulefill
\begin{eqnarray}\label{eqn:VarTCalc}
\begin{aligned}
\mathsf{Var}[|\mathcal{T}|] &= \mathsf{E}\left[\sum_{i \in \mathcal{I}} \mathbbm{1}(\hat{U}_i\neq U_i)\sum_{j \in \mathcal{I}} \mathbbm{1}(\hat{U}_j\neq U_j)\right]-\mathsf{E}\left[\sum_{i \in \mathcal{I}} \mathbbm{1}(\hat{U}_i\neq U_i)\right]\mathsf{E}\left[\sum_{j \in \mathcal{I}} \mathbbm{1}(\hat{U}_j\neq U_j)\right]\\
&=\frac{1}{2}\mathsf{E}\left[\sum_{i \in \mathcal{I}} \mathbbm{1}(\hat{U}_i\neq U_i)\sum_{j = i} \mathbbm{1}(\hat{U}_j\neq U_j)\right] + \frac{1}{2}\mathsf{E}\left[\sum_{i \in \mathcal{I}} \mathbbm{1}(\hat{U}_i\neq U_i)\sum_{j = i} \mathbbm{1}(\hat{U}_j\neq U_j)\right] \\
&\;\;\;+ \mathsf{E}\left[\sum_{i \in \mathcal{I}} \mathbbm{1}(\hat{U}_i\neq U_i)\sum_{j \neq i} \mathbbm{1}(\hat{U}_j\neq U_j)\right]- \mathsf{E}\left[\sum_{i \in \mathcal{I}} \mathbbm{1}(\hat{U}_i\neq U_i)\right]\mathsf{E}\left[\sum_{j \in \mathcal{I}} \mathbbm{1}(\hat{U}_j\neq U_j)\right]\\
&=\frac{1}{2}\mathsf{E}[|\mathcal{T}|]+\frac{1}{4}\mathsf{E}\left[\sum_{i \in \mathcal{I}} \mathbbm{1}(\hat{U}_i = ?)\sum_{j = i} \mathbbm{1}(\hat{U}_j = ?)\right] + \frac{1}{4}\mathsf{E}\left[\sum_{i \in \mathcal{I}} \mathbbm{1}(\hat{U}_i = ?)\sum_{j \neq i} \mathbbm{1}(\hat{U}_j=?)\right]\\
&\;\;\;- \frac{1}{4}\mathsf{E}\left[\sum_{i \in \mathcal{I}} \mathbbm{1}(\hat{U}_i=?)\right]\mathsf{E}\left[\sum_{j \in \mathcal{I}} \mathbbm{1}(\hat{U}_j=?)\right]\\
&=\frac{1}{2}\mathsf{E}[|\mathcal{T}|] + \frac{1}{4}\mathsf{E}\left[\sum_{i \in \mathcal{I}} \mathcal{E}_i\sum_{j \in \mathcal{I}} \mathcal{E}_j\right] -\frac{1}{4}\mathsf{E}\left[\sum_{i \in \mathcal{I}} \mathcal{E}_i\right]\mathsf{E}\left[\sum_{j \in \mathcal{I}} \mathcal{E}_j\right]\\
&=\frac{1}{2}\mathsf{E}[|\mathcal{T}|]+\frac{1}{4}\sum_{i \in \mathcal{I}}\sum_{j \in \mathcal{I}} \mathbf{C}_N(i,j).
\end{aligned}
\end{eqnarray}
\hrulefill
\end{figure*}





%\bibliographystyle{IEEEtran}
\bibliographystyle{IEEEtran}
\bibliography{Myreff2025}

\end{document}